\documentclass[10pt,aps,
notitlepage,%
oneside,%
twocolumn,
nofootinbib,%
superscriptaddress,%
floatfix, pra]%
{revtex4-2}
\pdfoutput=1
\usepackage[top=3cm, bottom=3cm, left = 2cm, right = 2cm]{geometry} 
\usepackage[utf8]{inputenc}
\usepackage{textcomp}
\usepackage{siunitx}
\usepackage[dvipsnames]{xcolor}
\usepackage{braket}
\usepackage{amsmath,amssymb}  
\usepackage{bm}  
\usepackage{mathtools}
\usepackage[pdftex,bookmarks,colorlinks,breaklinks]{hyperref} 
\usepackage{float}
\usepackage{graphicx,psfrag,rotating,epstopdf}% Include figure files
\usepackage{multirow}
\usepackage{tikz}
\usepackage[symbol]{footmisc}

\usepackage{ragged2e} % For text justification

\usepackage{physics}

\begin{document}
\title{Vector Magnetometry with Broadband Microwave Fields in Nitrogen-Vacancy Centers in Diamond}

\author{Tom~R.~Rieckmann*}
\affiliation{Institute for Physics, University of Rostock, Albert-Einstein-Straße 23-24, 18059 
Rostock, Germany}

\author{Arezoo~Afshar*}
\affiliation{National Research Council of Canada, 100 Sussex Drive, Ottawa, Ontario, K1N 5A2, Canada}
\affiliation{\mbox{Department of Physics, University of Ottawa, 25 Templeton Street, Ottawa, Ontario, K1N 6N5 Canada}}

\author{\\Aaron~Z.~Goldberg}
\affiliation{National Research Council of Canada, 100 Sussex Drive, Ottawa, Ontario, K1N 5A2, Canada}

\author{Lilian~Childress}
\affiliation{Department of Physics, McGill University, 3600 Rue University, Montréal, Québec H3A 2T8, Canada}

\author{Stefan~Scheel}
\affiliation{Institute for Physics, University of Rostock, Albert-Einstein-Straße 23-24, 18059 
Rostock, Germany}

\author{Khabat~Heshami}
\affiliation{National Research Council of Canada, 100 Sussex Drive, Ottawa, Ontario, K1N 5A2, Canada}
\affiliation{\mbox{Department of Physics, University of Ottawa, 25 Templeton Street, Ottawa, Ontario, K1N 6N5 Canada}}

\date{\today}
%
%============================================================
%Abstract
%============================================================
\begin{abstract}
    \noindent We present a novel method for full vector magnetometry using nitrogen-vacancy (NV) centers. In contrast to conventional optically detected magnetic resonance techniques, our method employs two distinct broadband microwave pulses and measures them after transmission through the NV sensor medium, thus capturing the line splitting of the ground state triplet due to the Zeeman effect. Two orthogonally polarized microwave pulses allow resolving all magnetic field components independently by reading out differently oriented NV centers. Simulated data is analyzed using deep neural networks, whose efficacy we expect to translate very well to experiments. Our method yields sensitivities between $5~\unit{\pico\tesla/\sqrt{\hertz}}$ and $100~\unit{\pico\tesla/\sqrt{\hertz}}$ across different magnetic field vector components, while achieving approximately $\unit{\nano\tesla}$ accuracy at a signal-to-noise (SNR) ratio of $70~\unit{\deci\bel}$. By being capable of accurately measuring magnetic fields down to $25~\unit{\micro\tesla}$, the need for a bias field beyond Earth's magnetic field is eliminated. 
\end{abstract}
\maketitle

\footnotetext[1]{Equal contribution. TRR developed the code for sampling, optimization, and data analysis. AA developed the simulation code used to obtain the time-domain signals using MW broadband addressing.}

%============================================================
\section{Introduction}
Beginning with the compass for navigation, measuring magnetic fields has repeatedly expanded our knowledge 
about the world and led to many novel technologies. Even today, the accurate determination of the entire 
magnetic field vector and its spatial variation results in discoveries and applications in physics, 
chemistry, biology, materials science, and 
engineering~\cite{Balasubramanian2008, s21165568, le2013optical, rondin2014magnetometry, nowodzinski2015nitrogen, Boto2018, PhysRevApplied.21.014055}. 
The development of more precise measurement techniques for different environmental settings is instrumental 
to these advancements. 

Nitrogen-vacancy (NV) centers have unlocked the possibility of highly precise magnetic field
estimation while embedded in a solid-state system at room temperature \cite{DOHERTY20131, Xu:23}. NV centers are point defects consisting of a substitutional nitrogen atom adjacent to a vacancy and have a symmetry axis oriented along one of the four bond directions in diamond.  Each NV orientation exhibits a distinct response to the same magnetic field, primarily determined by the field component along its symmetry axis.
% provides a unique 
% projection of the magnetic field vector $\vb{B}$. 
By probing the individual orientations, one can enhance 
sensitivity and reconstruct the full three-dimensional magnetic field vector~\cite{taylor2008high, Steinert2010ensemblesensitivity, clevenson2015broadband, chipaux2015magnetic, schloss2018simultaneous, munzhuber2017polarization}.
However, measurements typically require the magnetic field amplitude $\abs{\vb{B}}$ to be rather large, often necessitating a bias field of a few $\unit{\milli\tesla}$ to be added to the experimental setup~\cite{degen2017quantum}.  
\par %Add citation? Is this true for all readout methods?
We recently introduced a broadband microwave (MW) magnetometry method using NV centers in diamond, 
where a fast detector records the temporal evolution of the transmitted MW pulse and the magnitude of the magnetic field is estimated using the 
Kullback–Leibler (KL) divergence, achieving sub-$\mathrm{nT}/\sqrt{\mathrm{Hz}}$ 
sensitivity~\cite{afshar2025magnetometry}. The main advantages of this method are its reliable performance at 
magnetic fields as low as Earth's magnetic field ($25~\unit{\micro\tesla}$ to $65~\unit{\micro\tesla}$) without using a bias field and its ability to probe all NV 
orientations at once. \par 
As all 
NV orientations interact differently with the magnetic field, one naturally obtains information about the magnetic field's direction along with its strength. 
In this work, we extend our method in just this way, explicitly resolving $B_x$, $B_y$, and $B_z$. A single MW polarization is not sufficient to extract the full orientation~\cite{munzhuber2017polarization, mongelos2026unconditional}; hence, we make use of two MW polarizations to probe all orientations effectively. Additionally, we focus on analyzing the data 
with neural networks (NN) trained by machine learning (ML) instead of using the KL divergence. ML has
already been employed in a variety of systems for measurement 
analysis~\cite{doi:10.1126/sciadv.aat5218, PhysRevE.103.053312, 7tfd-22jr, SCHRODER2024113881}, including 
vector magnetometry \cite{Meng2023, Chen2022}. The main advantages for us are its straightforward 
implementation, its ability to avoid getting stuck in local minima, and its capability to accommodate 
different behaviors in experimental implementations, specifically in terms of the type of noise or artifacts 
in the data. This will allow experimental realizations to forego building accurate simulations of the 
physical process, including noise.  \par 

In Sec.~\ref{sec:Scheme}, we adapt the scheme developed in
Ref.~\cite{afshar2025magnetometry} to vector magnetometry and discuss the angular dependence. 
We then explain how we generated our data and performed the NN training in 
Sec.~\ref{sec:DataAndTraining}. Finally, Sec.~\ref{sec:Results} presents our results, including accuracies, 
precisions, and sensitivities that were obtained.

\section{Magnetometry Scheme}\label{sec:Scheme}
In our previous work, we developed a magnetometry technique based on a broadband microwave (MW) pulse 
interacting with an ensemble of NV centers in diamond, which enabled a measurement of the magnetic field magnitude without the 
need for frequency-domain scanning or optical readout~\cite{afshar2025magnetometry}. Instead of using 
conventional optically detected magnetic resonance (ODMR), which sweeps over many narrowband MW 
frequencies~\cite{zhu2023simulation}, a broadband MW pulse is employed to simultaneously address all 
magnetically sensitive spin transitions of the NV centers. 
Unlike conventional ODMR, where individual resonances are resolved sequentially, this approach encodes the entire resonance manifold into a single measurement.
As the MW pulse propagates through the NV ensemble, the spin-dependent absorption and dispersion modify the transmitted signal. As a result, the spectral response of the NV centers is encoded in the time-domain waveform of the transmitted MW field. By analyzing this time-domain signal, information about the external magnetic field can be extracted.

In order to accurately describe the NV center's interaction with the broadband field, we model the NV spin Hamiltonian, which includes electronic, nuclear, and hyperfine interactions. The Hamiltonian is given by
\begin{equation}
H = D S_z^2 + \gamma_e \mathbf{B} \cdot \mathbf{S}
+ \mathbf{S} \cdot \mathbf{A} \cdot \mathbf{I}
+ \gamma_n \mathbf{B} \cdot \mathbf{I},
\end{equation}
where $D$ is the zero-field splitting, $\gamma_e$ and $\gamma_n$ are the electronic and nuclear gyromagnetic ratios, respectively, and $\mathbf{A}$ is the hyperfine interaction tensor. The Zeeman term $\mathbf{B} \cdot \mathbf{S}$ governs the coupling between the magnetic field and the NV spin, leading to a response that depends on the orientation of the NV axis ($z$) relative to the applied field, and $\mathbf{I}$ is the nuclear spin vector.
 
From this Hamiltonian, we calculate the allowed transition frequencies for the ensemble of NV orientations. Each transition is modeled with a Lorentzian lineshape, with the linewidth set by the inhomogeneous dephasing time $T_2^*$. The individual transitions are summed over the NV orientations and hyperfine components to obtain the absorption strength. The overall absorption strength is scaled such that the total MW absorption through the diamond sample is approximately 16\%. Finally, we obtain the complex susceptibility $\chi(\omega) = \chi_1(\omega) + i \chi_2(\omega)$ using the Kramers-Kronig relations. The propagation of the MW field through the diamond sample with a length of $z$ and the refractive index $n$ is then written as
\begin{align} \label{eq:output_MW}
b_{MW}(z, \omega) = b_{0}(\omega) \exp\left[\frac{2\pi}{c}iz \left(n\omega + \omega_0 \frac{\chi(\omega)}{2}\right) 
\right],
\end{align}
where $b_{0}(\omega)$ is assumed to be a Gaussian input signal. The corresponding time-domain signal is obtained 
by applying the Fourier transform
\begin{equation}
b_{MW}(z,t) = \frac{1}{2\pi} \int_{-\infty}^\infty b_{MW}(z,\omega) e^{-i \omega t} 
\, d\omega.
\end{equation}
As the magnetic field shifts the NV transition frequencies, it changes the absorption profile and thus modifies the transmitted time-domain signal. Each magnetic field value produces a distinct temporal response. More details on how the time-domain signal is obtained are given in Appendix \ref{app:simulation_experimental_implementation}.

To enable full vector magnetometry, we consider two distinct linear polarizations of the MW drive, each represented by a polarization vector, 
to study how the coupling strength between the MW magnetic field and the NV spin depends on their relative 
orientation. By choosing appropriate MW polarizations, one can preferentially excite NV orientations whose symmetry axes are more strongly perpendicular to the microwave magnetic-field direction~\cite{kitazawa2017vector, yahata2019demonstration}.

\begin{figure}[htbp]
%    \centering
    \includegraphics[width=1\linewidth]{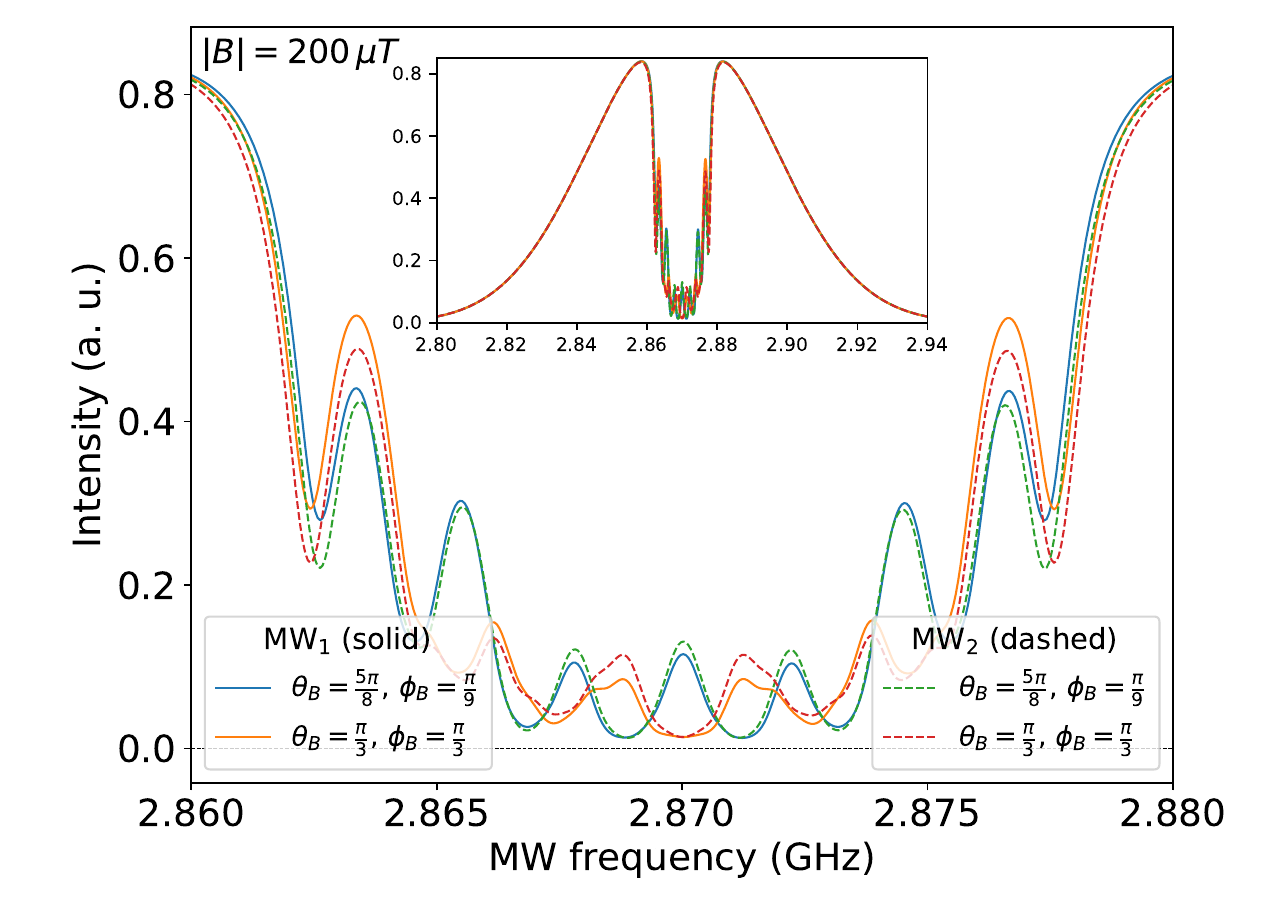}
        \caption{Frequency-domain response of the NV ensembles at a magnetic field of $|B| =$ 200$~\unit{\micro\tesla}$ for two orthogonal microwave polarization components. The first polarization vector, $\mathbf{b}_{\text{MW1}}$ (solid lines), is aligned along the $+z$-axis (\(\theta_1 = 0,\ \phi_1 = 0\)) with unit amplitude. The second polarization vector, $\mathbf{b}_{\text{MW2}}$ (dashed lines), lies in the transverse $xy$-plane with \(\theta_2 = \pi/2\) and \(\phi_2 = 7\pi/12\), also with unit amplitude. Each set of curves corresponds to a different field orientation \((\theta,\phi)\), illustrating the distinct spectral responses associated with the two linear MW polarization directions. The inset provides a zoomed-out view over the broader frequency range, illustrating the full resonance structure and relative peak amplitudes for both MW polarizations.} 
    \label{fig:signal_frequency_domain_two_MW}
\end{figure}

The laboratory coordinate system $(x,y,z)$ is chosen to coincide with the
$\langle 100\rangle$, $\langle 010\rangle$, and $\langle 001\rangle$
crystallographic directions, respectively. The polarization of the first
microwave magnetic field, $\mathbf{b}_{\mathrm{MW1}}$, is taken to be along
the laboratory-frame $+z$ axis, corresponding to $\theta_1=0$; the azimuthal
angle $\phi_1$ is arbitrary in this case and is set to zero for convenience, while the second, $\mathbf{b}_{\text{MW2}}$, lies in the transverse $xy$-plane with $\theta_{2} = \pi/2$ and $\phi_{2} = 7\pi/12$. An example of how the frequency-domain response changes with either the microwave polarization or magnetic field orientation is given in Figure \ref{fig:signal_frequency_domain_two_MW}.

\subsection{Angular Dependence}\label{subsec:AngularDependence}
The two MW pulses individually extract information about the magnetic field amplitude $\abs{\vb{B}}$ and partial information about its 
orientation. The orientation affects the measured signal mostly through two processes. First, the $m_s=\pm1$ energy levels are split 
due to the Zeeman effect, where the magnetic dipole $\vb*{\mu}$ interacts with the magnetic field $\vb{B}$ with a strength
proportional to $\vb*{\mu}\cdot\vb{B}$. The NV centers experience different shifts due to the four possible 
orientations of their magnetic dipoles. Second, the probing MW pulse interacts by magnetic dipole 
transitions in the NV centers, which are governed by the scalar product between the transition dipole and the 
MW polarization. In a simplified description, the first defines the spacing of peaks in the absorption 
spectrum, and the second defines the relative strength of peaks associated with different NV center orientations in 
the same spectrum. \par
Resulting from the way that NV centers contribute to the signal, we have found that, for any single microwave 
polarization, changes to the measured signal caused by adding a small $\delta \abs{\vb{B}}$ will mostly, if not 
completely, be canceled by small specific changes in the orientation $B_\phi$ and $B_\theta$. This renders an 
accurate reconstruction of $\vb{B}$ from a single MW pulse impossible at $\abs{\vb{B}}<50 ~\unit{\micro\tesla}$. For $\abs{\vb{B}}>50 ~\unit{\micro\tesla}$ the reconstruction is possible but still significantly less accurate. Hence, we always assume the use of two MW polarizations. More 
details for the behavior with a single MW polarization are found in Appendix \ref{app:singleMW}. \par 
Despite gaining information about the orientation of $\vb{B}$ using two microwave polarizations, symmetries 
caused by the Zeeman effect, such as when comparing $\vb{B}$ and $-\vb{B}$, are still unavoidable.
%With two microwave polarizations, although we can gain information about the orientation of $\vb{B}$, %symmetries caused by the Zeeman effect, such as when comparing $\vb{B}$ and $-\vb{B}$, are still
%unavoidable. 
Multiple $\vb{B}$ vectors will map to the same measured $I(t)$ and thus determining $\vb{B}$ from the 
signal $I(t)$ is not unique.
%does not have a unique solution. 

\begin{figure}[htb]
%    \centering
    \includegraphics[width=1\linewidth]{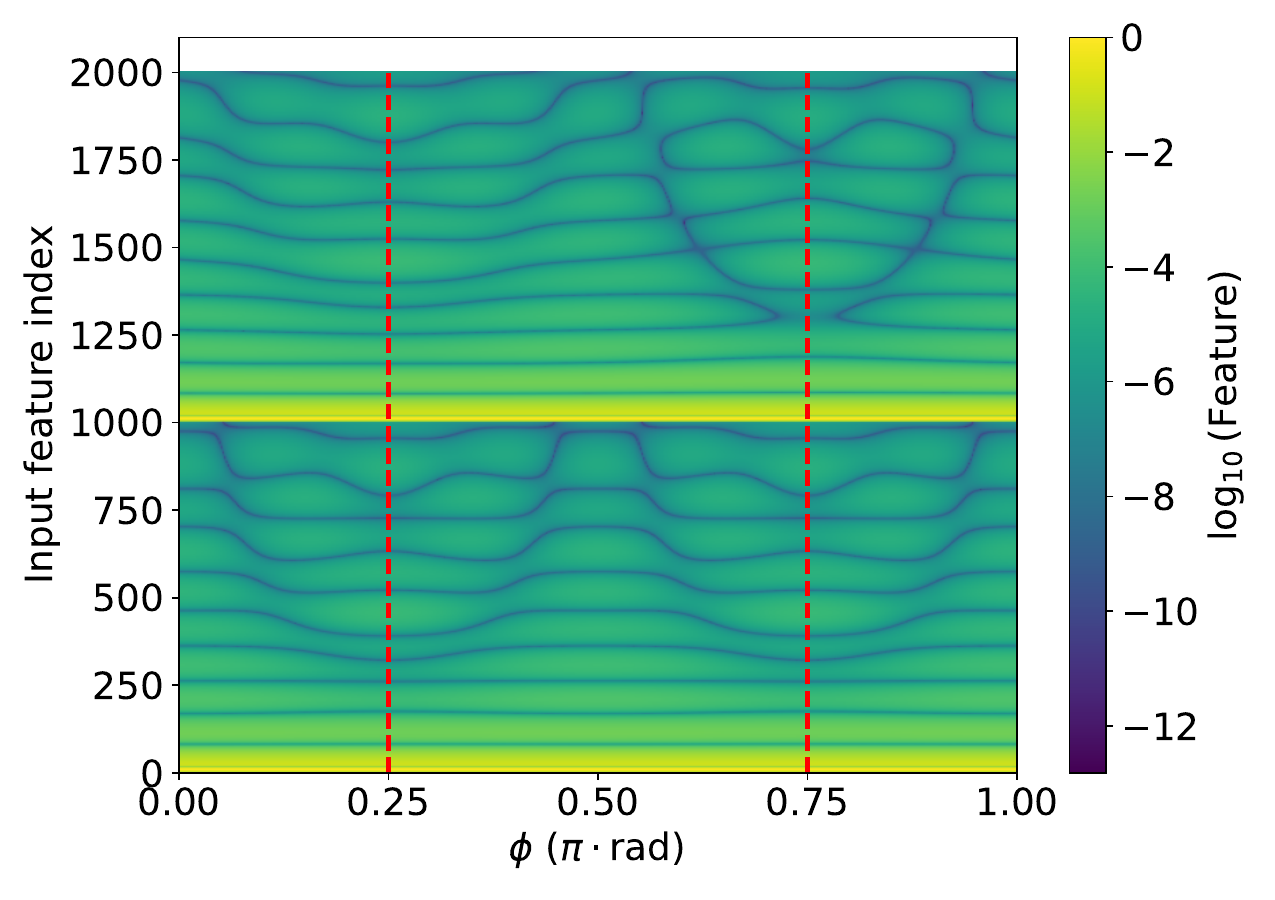}
    \caption{Concatenated signals $I_{\text{MW1}}(t)$ and $I_{\text{MW2}}(t)$  on the vertical axis after preprocessing for different $\phi$ with $\abs{\vb{B}}=100\unit{\micro\tesla}$ and $\theta = 3\pi/8$. The red dashed lines show symmetry axes for the local behavior around them.} 
    \label{fig:PhiDependenceSymmetry}
\end{figure}

% Second figure left out as it doesnt really give any additional info
% \begin{figure}
%     \centering
%     \includegraphics[width=1\linewidth]{Images/theta_dependence_input_wide.pdf}
%     \caption{Concatenated signals $I_{\text{MW1}}(t)$ and $I_{\text{MW2}}(t)$ after preprocessing  on the vertical axis for different $\theta$ with $\abs{\vb{B}}=100\unit{\micro\tesla}$ and $\phi = \pi/2$. The red dashed lines show symmetry axes for the local behavior around them.} 
%     \label{fig:ThetaDependenceSymmetry}
% \end{figure}

Figure~\ref{fig:PhiDependenceSymmetry} exemplifies the resulting symmetry boundaries $\phi_{\text{sym}}$ at $\{\frac{\pi}{4}, \frac{3\pi}{4}\}$, i.e. the two signals for $\phi_{\text{sym}} + \Delta \phi$ and $\phi_{\text{sym}} - \Delta \phi$ are identical.
For the orientation of NV centers in our chosen frame and the selected microwave polarizations, we have observed 
that the unavoidable symmetry boundaries lie parallel to the axes of our spherical coordinate system. 
Specifically, we found empirically that if the allowed $\theta$ values in the dataset cross $\frac{\pi}{4};~\frac{\pi}{2};~\frac{3\pi}{4}$, or values of $\phi$ cross 
$\frac{\pi}{4};~\frac{3\pi}{4};~\frac{5\pi}{4};~\frac{7\pi}{4}$, the signal for different points is not unique and identifying an accurate 
estimate of $\vb{B}$ becomes impossible. Adding more MW polarizations could eliminate these boundaries, leaving 
only the symmetry between $\vb{B}$ and $-\vb{B}$. However, this would come with significant additional 
experimental requirements and still does not solve the symmetry problem completely. Moreover, we expect the 
sensitivity would be strongly dependent on the orientation of the measured $\vb{B}$.
Instead, if a guess for $\vb{B}$ is known in advance, one can orient the crystal for the measurement appropriately to avoid these boundaries. If 
there is no information about $\vb{B}$ available, we propose to solve the symmetry problem by combining our sensor with another less accurate sensor, 
which only gives a rough estimate of the magnetic field orientation. The crystal should then be reoriented so 
that $\vb{B}$ lies within our desired region in the sensor's reference frame. For further analysis, we have 
limited our sampling to one region avoiding symmetry boundaries, with details given in 
Sec.~\ref{subsec:SampleGen}. 

% \begin{figure}[htb]
%     \centering
%     \includegraphics[width=1\linewidth]{Images/symmetry_boundaries_insphere.pdf}
%     \caption{Boundaries that a magnetic field vector sampling region may not cross to ensure that the function associating the magnetic field vector to the microwave signal is invertible. Otherwise, the problem of selecting the correct magnetic field vector does not have a unique solution. Boundaries for $\theta$ are at $\pi/4, \pi/2, 3\pi/4$, those for $\phi$ at $\pi/4, 3\pi/4, 5\pi/4, 7\pi/4$.} 
%     \label{fig:SamplingBoundaries}
% \end{figure}

Although such a reorientation procedure introduces additional experimental complexity, we prefer it to
alternative solutions. For methods that try to 
break the symmetry by providing information from the additional sensor only in postprocessing, it is impossible 
to avoid the noisy estimate landing on the wrong side if the initial $\vb{B}$ is too close to a symmetry 
boundary, which would give highly inaccurate results for the NV-based sensor. Moreover, limiting the region of 
interest has an additional advantage in allowing us to increase the density of training data samples.

\begin{figure*}
    \centering
    \includegraphics[width=0.9\linewidth]{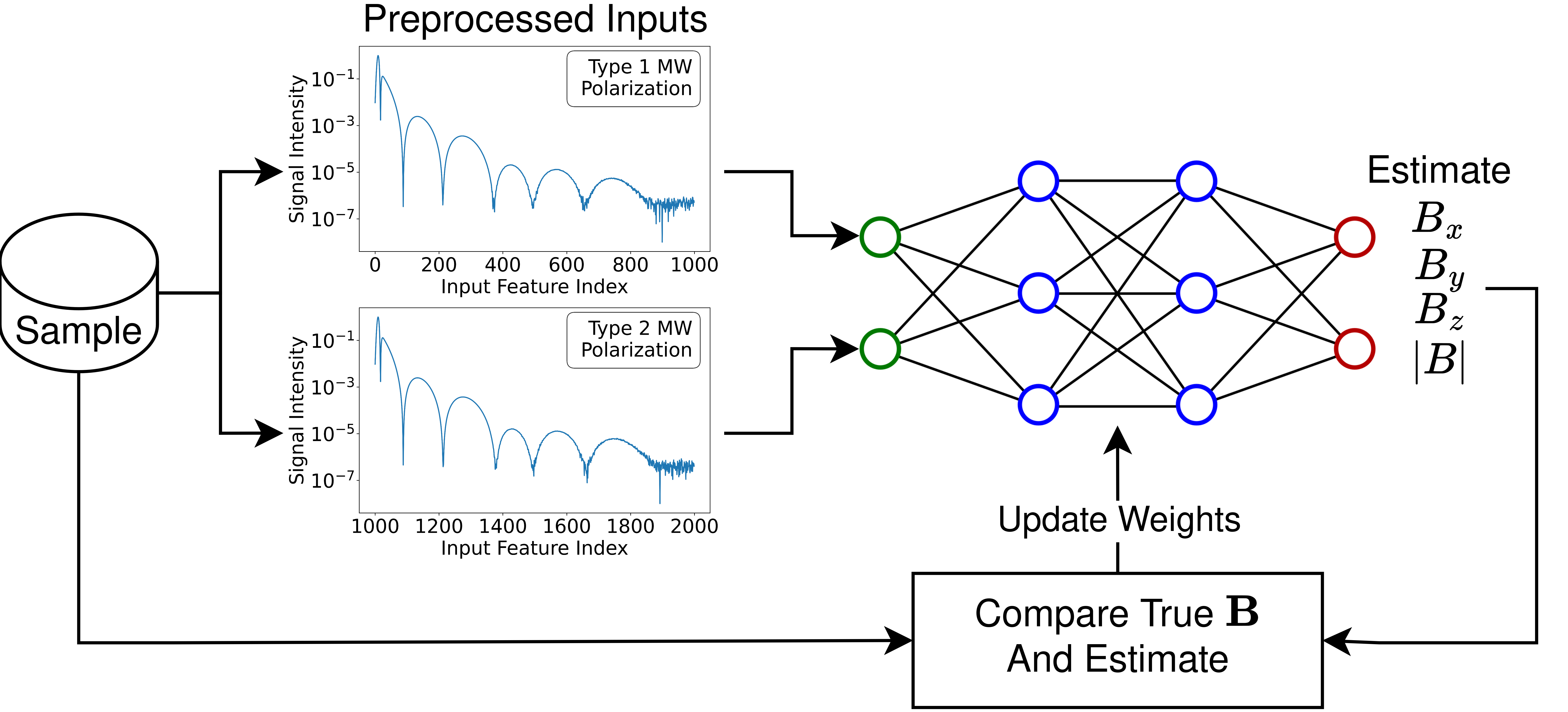}
    \caption{Depiction of NN-based Vector Magnetometry. Apart from a few simple preprocessing steps, the NN is tasked with directly giving $B_x, B_y, B_z, \lvert B\rvert$ as output. The weights of the NN are updated during training using the mean square error between the true and estimated $\vb{B}$.} 
    \label{fig:NNTraining}
\end{figure*}

\section{Neural Network-based Estimation} \label{sec:DataAndTraining}
As discussed above, we are able to simulate the resulting measurable intensities 
$I(t) = \{I_{\text{MW1}}(t), I_{\text{MW2}}(t)\}$ for a finite set of selected magnetic fields $\vb{B}$. 
We solve the inverse problem of determining $\vb{B}$ from $I(t)$ by training a deep neural network (NN) on a set of simulated samples $\{I_k(t), \vb{B}_k\}$. \par

\subsection{Sample Generation}\label{subsec:SampleGen}
Samples $\vb{B}_k$ are generated with uniform density within a specified volume. We then calculate the fields after passing through the sensor crystal (see Appendix~\ref{app:simulation_experimental_implementation}) for each sample. Noise following a normal distribution is added to training, validation, and test data, matching signal-to-noise ratios (SNRs) that one would expect in experiments. $I_{\text{MW1},k}(t)$ and $I_{\text{MW2},k}(t)$ are obtained from the noisy electric fields. To match the simulation, a sampling rate of $1~\unit{\giga\hertz}$ is assumed. 
Reference~\cite{afshar2025magnetometry} discusses the possibility of using different sampling rates and shows how the sensitivities change with their rates. \par 
For our main results, we have elected to bound the $\vb{B}$-values by sampling from a cylinder in $\vb{B}$-space 
of radius $r=2~\unit{\micro\tesla}$ 
placed along the radial component. A second approach with a less universally applicable, but simpler 
experimental implementation, is given in Appendix~\ref{app:MethodB}. The cylinder has a sufficiently large radius that 
inaccuracies of an initial sensor's rough estimation will not cause the real data points to be outside of it. 
In this case, we can freely select the range of $\abs{\vb{B}}$ that is of interest, so long as one avoids the 
symmetry boundaries for weak magnetic fields. We have chosen a range 
$\abs{\vb{B}} \in [20~\unit{\micro\tesla}, 100~\unit{\micro\tesla}]$ with $\theta_{\text{cyl}} = \frac{3}{8}\pi$ and $\phi_{\text{cyl}} = \frac{\pi}{2}$. 
The given upper limit for $\abs{\vb{B}}$ is purely suggested for the analysis in this paper and 
can be adjusted based on experimental requirements and limitations.
The experimental method would consist of making an initial estimate with the other sensor, e.g. a low-cost vector magnetometer such as \cite{akm_ak09940a_Magnetometer}, then 
rotating the NV-based sensor so that this initial estimate is on the cylinder's central axis, and at last 
obtaining a highly refined estimate from the NV-based sensor. This method is applicable with no further 
constraints beyond the range of $\abs{\vb{B}}$, as the reorientation for each new measurement eliminates the 
limitations caused by symmetry boundaries.

\subsection{Neural Network Structure and Training}

In order to obtain $\vb{B}$ from $I_{\text{MW1}}(t)$ and $I_{\text{MW2}}(t)$, we train a NN following 
standard procedures in supervised learning with deep neural networks \cite{Bishop2023}. A schematic is given 
in Figure~\ref{fig:NNTraining}. For each sample $k$, we first perform some preprocessing steps. The signal is reduced to the part containing the MW pulse with the material response, i.e. a time window of $1~\unit{\micro\second}$. This creates two
vectors $I_{1,k}, I_{2,k} \in \mathbb{R}^{d_\text{in}}$, with $d_\text{in}=1000$, though this dimension could vary depending on the sampling rate and the noise floor. We then obtain our NN input vector as
\begin{equation}
    \vb{V}_{\text{in}} = \pmqty{I_{1,k}/\norm{I_{1,k}}_\infty \\ I_{2,k}/\norm{I_{2,k}}_\infty }.
\end{equation}
The output vectors are used directly as estimates $B_{i,\text{est}}: B_x, B_y, B_z, B_\text{abs}$, where 
$B_\text{abs}$ is included for convenience. We equate $B_\text{abs}=\abs{\vb{B}}$, and any quantity with index $i$ includes the fourth parameter, i.e $i \in \{x,y,z,\text{abs}\}$. Adding $\abs{\vb{B}}$ to the estimated output does not yield more information, as it is generated from the same input, and deviations in the estimation of parameters are correlated. More details are given in Appendix \ref{app:Overdetermined}. Additional steps, such as treating the output as an estimate in 
spherical coordinates or scaling it to the range $[0, 1]$, did not give significantly better results in our 
experience. However, such steps could become necessary if one were to use different bounds for the sampling 
regions.  NN weights are updated during training using the ADAM optimizer 
\cite{kingma2017adammethodstochasticoptimization} on the mean square error (MSE) loss
\begin{equation}
    L = \frac{1}{4 \lvert D\rvert} \sum_k^D \sum_i (B_{i,\text{true}} - B_{i,\text{est}})^2
\end{equation}
with $i = \{x, y, z, \text{abs}\}$. $D$ represents a mini-batch of the total training set. \par 
We use a feedforward NN to perform the estimation due to its simplicity and the fact that, when we tested 
other structures, they performed at best similarly while requiring longer training times. The exact training 
parameters were determined using random search hyperparameter optimization and can be found in 
Appendix~\ref{app:NNParams}.
\par 
After training, there may still exist a remaining estimation error in the NN-based estimation, which we call $\Delta B$. 
Even if we ignore all noise to get ideal simulated inputs $\vb{V}_{\text{in,ideal}}$ (or we average over infinitely
many noisy measurements), the output still contains a bias of the form
\begin{equation}
    \vb{B}_{\text{est,ideal}} = f_{NN}(\vb{V}_{\text{in,ideal}}) = \vb{B}_{\text{true}} + \Delta \vb{B}.
\end{equation}
As such, our approach acts as a biased estimator, with $\Delta\vb{B}$ depending on $\vb{B}_{\text{true}}$ 
that we try to estimate, or rather the corresponding $\vb{V}_{\text{in,ideal}}$ in an ideal system. 
Fortunately, it is possible to reduce the bias so far that it becomes small compared to the effect 
of noise at realistic values for the SNR. Details on our techniques to achieve this are given in 
Appendix~\ref{app:bias}.

\section{Results}\label{sec:Results}
\begin{table*}[]
\setlength{\tabcolsep}{3pt}
\renewcommand{\arraystretch}{1.4}
\begin{tabular}{c|cc|cc|cc|cc|}
\cline{2-9}
 & \multicolumn{2}{c|}{$B_x$}                & \multicolumn{2}{c|}{$B_y$}                & \multicolumn{2}{c|}{$B_z$}                & \multicolumn{2}{c|}{$\abs{\vb{B}}$}              \\ \hline
\multicolumn{1}{|c|}{}       & \multicolumn{1}{c|}{$\sigma_{x, \text{est}}~(\unit{\nano\tesla})$} & $\eta_x~(\unit{\pico\tesla/\sqrt{\hertz}})$ & \multicolumn{1}{c|}{$\sigma_{y, \text{est}}~(\unit{\nano\tesla})$} & $\eta_y~(\unit{\pico\tesla/\sqrt{\hertz}})$ & \multicolumn{1}{c|}{$\sigma_{z, \text{est}}~(\unit{\nano\tesla})$} & $\eta_z~(\unit{\pico\tesla/\sqrt{\hertz}})$ & \multicolumn{1}{c|}{$\sigma_{\text{abs,est}}~(\unit{\nano\tesla})$} & $\eta_\text{abs}~(\unit{\pico\tesla/\sqrt{\hertz}})$ \\ \hline
\multicolumn{1}{|c|}{Min}    & \multicolumn{1}{c|}{7.2(2)} & 72 & \multicolumn{1}{c|}{1.31(4)} & 13            & \multicolumn{1}{c|}{0.60(2)}    & 6.0 & \multicolumn{1}{c|}{1.26(4)}    &    13         \\ 
\multicolumn{1}{|c|}{Median} & \multicolumn{1}{c|}{8.3(3)}    &  83         & \multicolumn{1}{c|}{1.56(5)}    &   16         & \multicolumn{1}{c|}{0.97(3)}    &    9.7        & \multicolumn{1}{c|}{1.49(5)}    &    15       \\ 
\multicolumn{1}{|c|}{Max}    & \multicolumn{1}{c|}{11.8(4)}    &     118       & \multicolumn{1}{c|}{3.2(1)}    &     32        & \multicolumn{1}{c|}{4.8(2)}    &      48       & \multicolumn{1}{c|}{2.17(7)}    &     22      \\ \hline
\end{tabular}
\caption{Select standard deviations $\sigma_{x, \text{est}}$ and sensitivities $\eta_i$ from a set of 100 test data points are given for all four 
estimated parameters at an SNR of $70~\unit{\deci\bel}$ with an estimated measurement time of 
$100~\unit{\micro\second}$. Values inside brackets are the uncertainty in the estimation of the standard deviation.}
\label{tab:StdAndSensitivity}
\end{table*}

\begin{figure}
    \centering
    \includegraphics[width=1\linewidth]{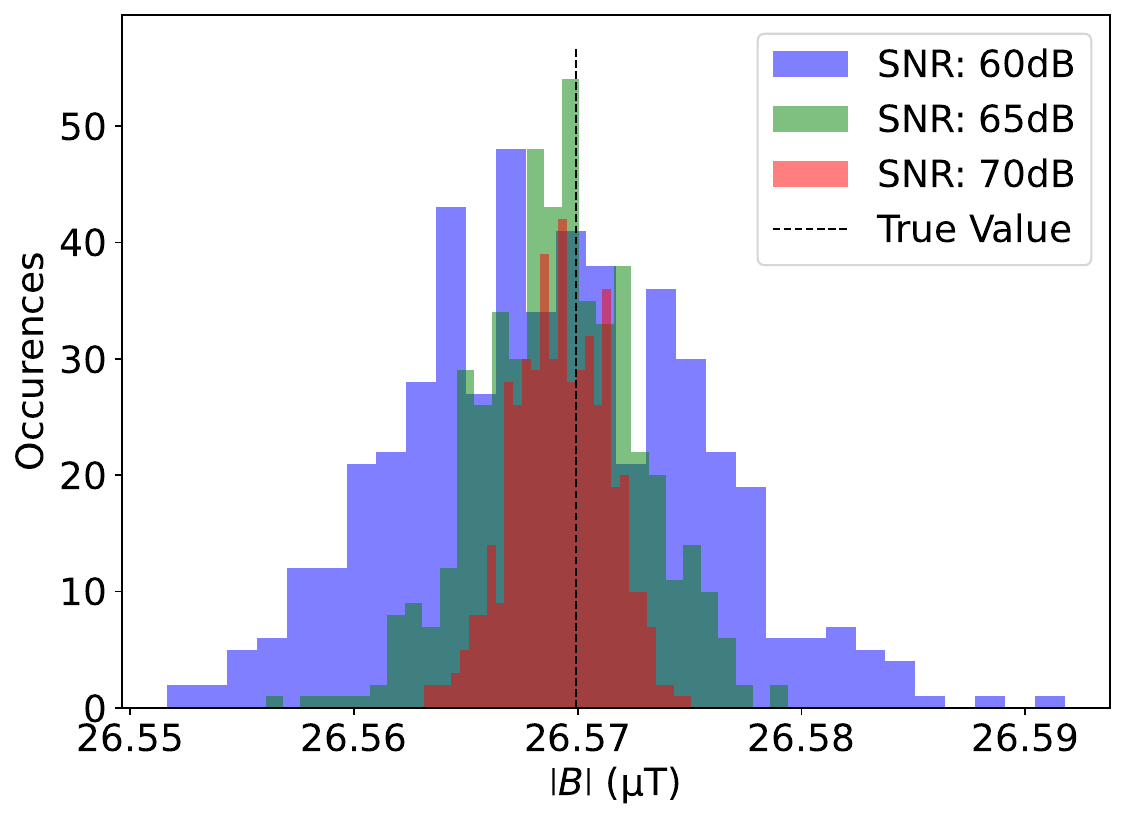}
    \caption{Example histogram from the estimation of $\abs{\vb{B}}$ using 500 different noisy samples at each SNR.} 
    \label{fig:HistogramExample}
\end{figure}
Training and validation were performed with 100000 individual data points, i.e. each $\vb{B}_{\text{true}}$ was 
associated with a single noisy $\vb{V}_{\text{in}}$. To analyze the statistics of the parameter 
estimation, we created a separate test dataset consisting of 100 individual magnetic field vectors sampled 
within the specified volume, independently of the training and validation dataset. Then, for each data point $\vb{B}_{\text{true}}$
and at each tested SNR, we sampled 500 different random noisy profiles to be added to the ideal signal, creating 500 instances $\vb{V}_{\text{in}}$ each. 
As the noise propagates through the NN layers, we end up with distributions for the estimated parameters. 
An example is shown in Figure~\ref{fig:HistogramExample}. For the other estimated parameters and for different 
$\vb{B}_{\text{true}}$ one obtains similar histograms. For simplicity, all further analysis is performed using the obtained means and standard deviations for each $\vb{B}_{\text{true}}$. If the noise amplitude is increased and the data 
point lies close to the surface of the sampled volume, the distribution may become skewed as the NN learns in 
training that no data point lies outside the sampling region. \par 
\begin{figure}
    \centering
    \includegraphics[width=1\linewidth]{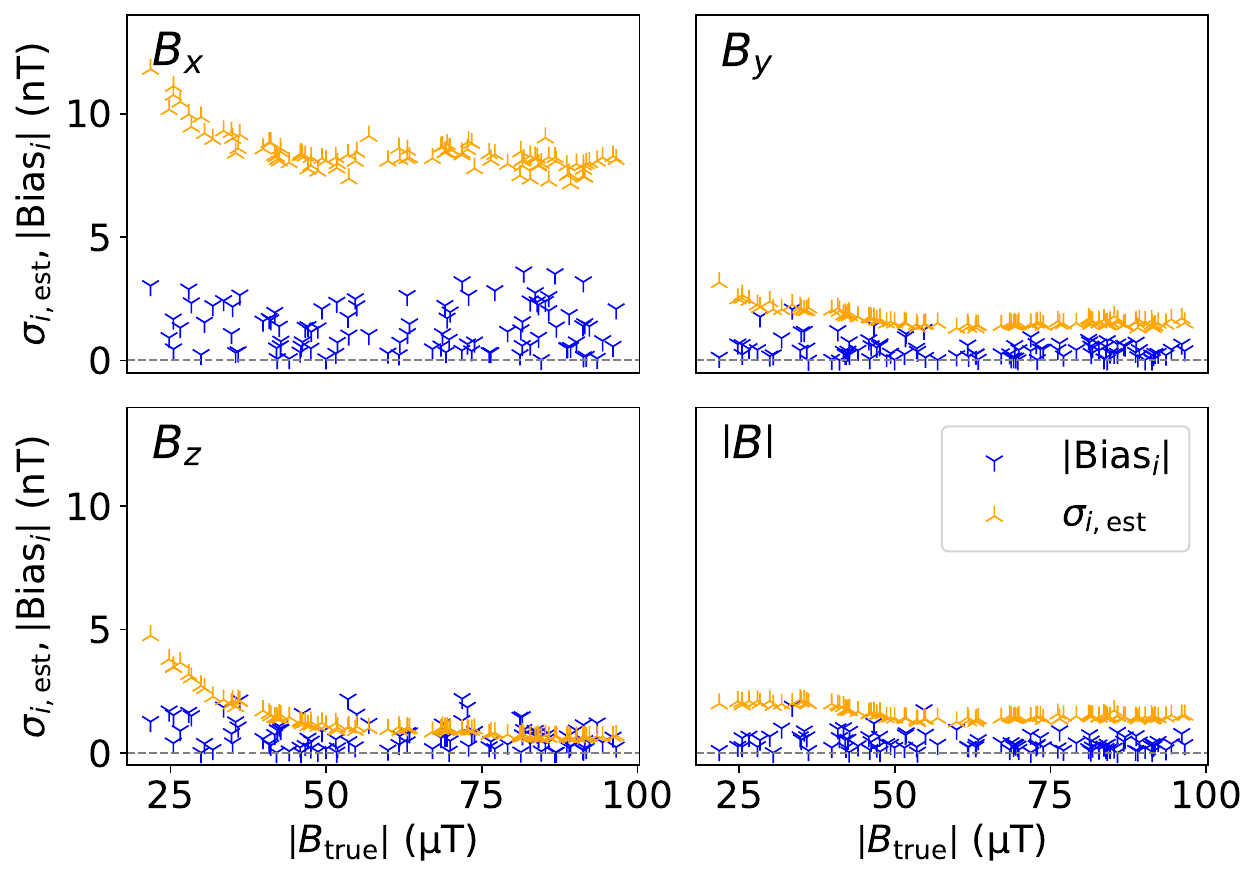}
    \caption{$\sigma_{i, \text{est}}$ and Bias of the estimation for all $B_i$ with an SNR of $70~\unit{\deci\bel}$ in Method A. Each data point is extracted from a distribution of a series of estimations at the same $\vb{B}_{\text{true}}$, as depicted in Figure~\ref{fig:HistogramExample}. One can see $\sigma_{i, \text{est}}$ strongly depends on $\abs{\vb{B}_{\text{true}}}$, whereas the Bias is mostly independent. Values for the minimum, maximum, and median of $\sigma_{i, \text{est}}$ are given in Table~\ref{tab:StdAndSensitivity}. } 
    \label{fig:BiasSTDCylinderSmaller}
\end{figure}
To simplify our findings, the two main parameters from the estimation that we are interested in are the 
standard deviation $\sigma_{i, \text{est}}$ for each parameter $B_i$ and the absolute value of the bias
\begin{equation}
    \abs{\text{Bias}_i(B_{i, \text{est}}, B_{i, \text{true}})} =  \abs{\mu_{i, \text{est}} - B_{i, \text{true}}},
\end{equation}
where $\mu_{i, \text{est}}$ is the mean of the obtained estimates for a singular $B_i$ in the test dataset. Details about the properties of the bias can be found in Appendix~\ref{app:bias}. Upper and lower bounds for the 
obtained standard deviations in the test dataset are given in Table~\ref{tab:StdAndSensitivity}. 
Figure~\ref{fig:BiasSTDCylinderSmaller} shows the dependence of these parameters on $\abs{\vb{B}_{\text{true}}}$ 
over the 100 test data points. The bias mostly strays randomly and is at or below the 95\% confidence level 
obtained from $\sigma_{i, \text{est}}$. One can see that the standard deviation for $\abs{\vb{B}}$ is almost 
constant. For  $B_x$, $B_y$, and $B_z$ it decreases asymptotically as $\abs{\vb{B}_{\text{true}}}$ increases. We 
attribute this behavior to how the NN determines its estimate. For low $\abs{\vb{B}_{\text{true}}}$, the change in the signal for small variations $\delta\vb{B_{\text{true}}}$ along the $x$, $y$, and $z$-axes is smaller. The noise has an approximately constant amplitude over the dataset. Thus, it affects the estimate more strongly for small $\abs{\vb{B}_{\text{true}}}$, which causes an increase in 
$\sigma_{i, \text{est}}$. $\sigma_{x, \text{est}}$ is larger than the others, because---for our sampling region and MW polarizations---$B_x$ has the smallest effect on the signal. If one changes the MW polarizations or sampling region, the performance along other axes will be degraded, offsetting the improvements in $B_x$. An example of how the change in the signal is axis-dependent is given in Figure \ref{fig:KLDisappereance1MW}. The higher precision of the $\abs{\vb{B}}$ estimation cannot be used to improve the precision of $B_x$, since the axes of these two parameters are effectively perpendicular within our sampling region (see Appendix \ref{app:Overdetermined}). \par 

To give an estimation of the sensitivity $\eta_i$, we will now focus on the best case, worst case, and median 
performance of our estimator on the test data points. The corresponding results can be seen in 
Table~\ref{tab:StdAndSensitivity}. For the calculation of the sensitivity $\eta_i$, we use the same 
approximate measurement duration of $T=100~\unit{\micro\second}$ for a single pulse as in 
Ref.~\cite{afshar2025magnetometry}.
The single-shot SNR was estimated to be around $75~\unit{\deci\bel}$ in the previous work; we will use 
$\mathrm{SNR}_\mathrm{init}= 70~\unit{\deci\bel}$ for our calculation. The sensitivities were obtained from
\begin{equation}
    \eta_i = \sigma_{i,\text{est}} \sqrt{T}
\end{equation}
and are on the order of tens of $\unit{\pico\tesla/\sqrt{\hertz}}$.

\section{Conclusion}
In this work, we have extended the previously introduced method for determining the magnetic field strength 
using a single broadband microwave pulse. By adding a second microwave pulse with a different polarization, 
one can instead determine the full magnetic field vector. Analysis of the measurements was performed using 
machine learning. With this method we predict sensitivities of  $10\unit{\pico\tesla/\sqrt{\hertz}}$ to 
$100\unit{\pico\tesla/\sqrt{\hertz}}$ for any vector component in experimental realizations under the 
assumption that $\abs{\vb{B}}$ is within the range of Earth's magnetic field. With the NN trained for this paper, an accuracy
of around $1~\unit{\nano\tesla}$ is achievable, although this value can be decreased by dedicating more resources
to NN training or by combining the ML estimation with an unbiased estimator (e.g. KL divergence in Ref.~\cite{afshar2025magnetometry}) that may use the NN estimate as its initial value, thus bypassing runtime restrictions.
\par 
For an experimental realization, we have presented a general approach applicable down to 
$25~\unit{\micro\tesla}$, necessitating only the reorientation of the measurement device. However, a more 
simplified implementation is discussed in Appendix~\ref{app:MethodB}, which just requires that $\vb{B}$ only 
varies up to $20~\unit{\micro\tesla}$ around a known background, such as the local magnetic field of the Earth. 
While the method from Appendix~\ref{app:MethodB} does decrease the maximal accuracy, this reduction can be offset if $\vb{B}$ is known to deviate less or by dedicating more resources to data gathering. \par

We believe that our method will allow NV sensors to be applied in cases where they were so far quite limited in their usefulness. There is no need for a bias magnetic field and the only steps to the estimation, apart from the measurement itself, are either limiting the setup to a fixed background, or gathering an initial guess from a small, commercially available magnetometer and implementing a rotation mechanism. We have shown it to be applicable while only in the presence of the background field of the Earth and it retains other advantages of NV sensors, such as their small sensor size and room-temperature operability.

\section*{Acknowledgements}
TRR and SS acknowledge financial support by the Deutsche Forschungsgemeinschaft via the International Research Training Group 2676 `Imaging quantum systems (IQS): photons, molecules, materials’, grant no. 437567992.
AA, AZG, and KH acknowledge that the NRC headquarters is located on the traditional unceded territory of the Algonquin Anishinaabe and Mohawk people. All authors thank SBQuantum and Noah Lupu-Gladstein for fruitful discussions. KH acknowledges funding from the NSERC Discovery Grant. AA, AZG, LC, KH, and SS acknowledge funding from NRC's Quantum Sensors Program.

\bibliography{references}

\appendix
\section{Simulation Parameters and Experimental Considerations for Broadband MW Polarization 
Addressing}\label{app:simulation_experimental_implementation}
Here, we summarize the simulation and experimental assumptions underlying the broadband MW magnetometry 
scheme described in the main text, following the method developed in Ref.~\cite{afshar2025magnetometry}.
\subsection{Broadband MW Input Pulse}
The incident MW field is modeled as a Gaussian pulse in the frequency domain, 
\begin{align} \label{eq:input_MW}
b_{0}(\omega) = \frac{1}{\sigma\sqrt{2\pi}}\,\text{exp}\Big[-\,\frac{(\omega-\omega_0)^2}{2\sigma^2}\Big],
\end{align}
centered at the NV zero-field splitting of $\omega_0=2.87~\unit{\giga\hertz}$.
In all simulations, the pulse bandwidth is fixed to $\sigma = 85~\unit{\mega\hertz}$. This bandwidth is sufficient to simultaneously 
address all magnetically sensitive spin transitions of the NV ensemble over the considered magnetic field range. 
Moreover, the temporal signal is calculated by Fourier transforming the transmitted microwave field from the frequency domain to the time domain. The resulting time-domain field is sampled with $1~\unit{\nano\second}$ temporal resolution over a $1~\unit{\micro\second}$ time window. To model detector/electronic noise, additive noise is applied to
the time-domain field amplitude before calculating the intensity. The recorded
intensity signal is then obtained as
\[
I(t)\propto |b_{MW}(z,t)|^2,
\]
where $b_{MW}(z,t)$ denotes the transmitted microwave magnetic-field
amplitude. The 1~$\mu$s time window captures the region of highest sensitivity to
changes in the external magnetic field.

\subsection{Polarization Control}
Experimentally, the broadband MW field pulse can be delivered to the diamond via a $50~\unit{\ohm}$ stripline 
waveguide, producing a MW magnetic field that couples to the NV spins. Two distinct linear MW polarizations can 
be implemented using two orthogonal striplines. 
To distinguish the scalar spectral amplitude of the pulse from its polarization direction, we write the frequency-domain MW field as
\begin{align}
    \mathbf{b}_{\mathrm{MW},j}(\omega)
    =
  b_0(\omega)\,\hat{\mathbf{e}}_{\mathrm{MW},j},
    \qquad j=1,2,
\end{align}
where $b_0(\omega)$ is the scalar spectral envelop defined in Eq.~\eqref{eq:input_MW}, and \(\hat{\mathbf{e}}_{\mathrm{MW},j}\) is a unit vector specifying the MW polarization direction.
The two polarization direction considered in this work are:
\begin{align*}
    \hat{\mathbf{e}}_{\mathrm{MW},1}: 
    &\quad (\theta_1 = 0, \,\phi_1 =0), \\
    \hat{\mathbf{e}}_{\mathrm{MW},2}: 
    &\quad (\theta_2 = \pi/2, \,\phi_2 = 7\pi/12),
\end{align*}
each with unit amplitude. These polarization directions weigh the four crystallographic NV orientations differently, 
as only the component of the MW magnetic field transverse to a given NV symmetry axis drives the 
\(|m_s = 0\rangle\longleftrightarrow|m_s = \pm1\rangle\) transitions. Consequently, the relative strengths of the absorption features in the ensemble susceptibility depend on the chosen MW polarization, leading to 
distinct time-domain signals for the two cases.

\subsection{NV Density and MW Absorption}
The simulations assume a dense NV ensemble representative of commercially available diamonds such as E6 B14, 
with a total NV concentration of approximately 4.5 ppm. Accounting for realistic experimental conditions, namely,
$\sim70\%$ occupancy of the negatively charged NV$^-$ state and $\sim80\%$ optical polarization into the 
$|m_s=0\rangle$ state, the effective polarized NV density is taken to be 
\(n_{\mathrm{NV}}^{\mathrm{eff}} \simeq 0.56~\mathrm{ppm}\) per orientation.

Under weak MW drive conditions (negligible power broadening), which is necessary for the linear response 
function, and assuming an inhomogeneous dephasing time $T_2^* = 1~\unit{\micro\second}$, the microwave absorption 
coefficient is estimated to be $\sim1.28~\unit{\deci\bel/\centi\meter}$, corresponding to approximately 25\% MW power 
absorption per centimeter of propagation through the NV-doped region. In the simulations presented here, a 
total absorption of $\sim16\%$ is assumed to demonstrate the effect of NV-induced 
absorption and dispersion on the transmitted broadband MW pulse.

\subsection{Noise Model}
The dominant noise source in the MW detection is assumed to be Johnson-Nyquist noise~\cite{eisenach2021cavity}. 
Over a detection bandwidth matching the pulse bandwidth \(\Delta f\approx 85~\unit{\mega\hertz}\), the root-mean-square noise voltage at room temperature is
\[V_{\mathrm{JN}}=\sqrt{4k_B T R \Delta f}\approx 8.4~\unit{\micro\volt},\]
for $T=300~\unit{\kelvin}$ and $R=50~\unit{\ohm}$. For the detected MW signal the rms voltage of $V_{rms} \approx 50~\unit{\milli\volt}$, which corresponds 
to a SNR of approximately $75~\unit{\deci\bel}$ relative to the thermal noise. 

\section{Estimation with one MW}\label{app:singleMW}
\begin{figure}
    \centering
    \includegraphics[width=1\linewidth]{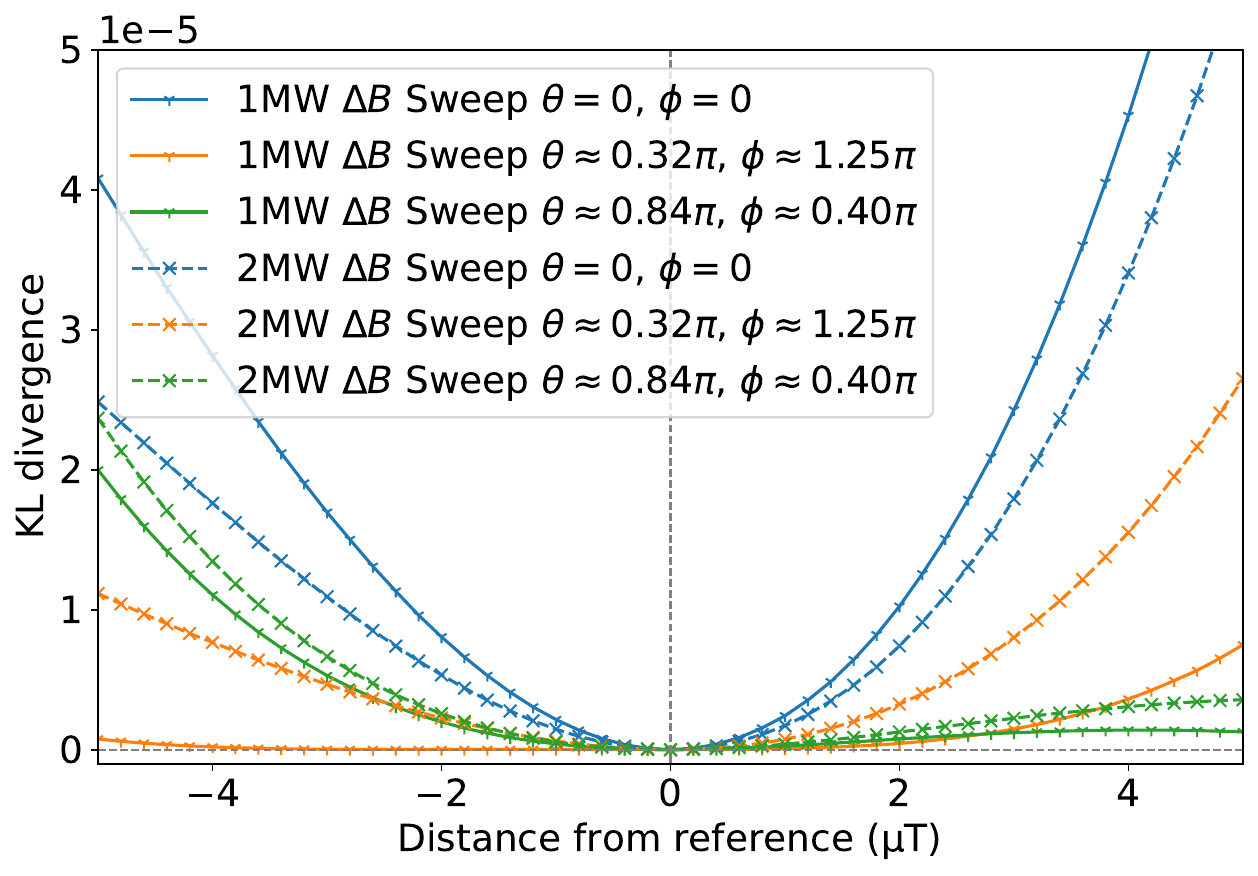}
    \caption{Change of KL divergence when treating the simulated signals as probability distributions after normalization. Starting from a fixed reference positioned at $\vb{B} \approx (-16.3, 16.3, 9.57)~\unit{\micro\tesla}$, the curves are created by walking the same distance along different sweep directions (see legend, values were rounded). For a single MW polarization (solid lines, $\theta_B=\frac{\pi}{3}, \phi_B = 0$), we could find sweep directions in which there is practically no change in the measured signal (see orange plot), or the change flattens out and decreases again (see green plot). Adding a second MW polarization ($\theta_B=\frac{\pi}{2}, \phi_B = \frac{\pi}{2}$) can resolve this issue (dashed lines).}
    \label{fig:KLDisappereance1MW}
\end{figure}

We have extensively tested to see if the experimental protocol can be made to work with a single MW polarization, while not detracting from the usable region of magnetic fields. While symmetries in the generated signal along the usual axes ($B_x, B_y, B_z, B_r, B_\theta, B_\phi$) are avoidable, the performance is still rendered significantly worse, even when limiting the sampling regions to a radius of only $2~\unit{\micro\tesla}$. Specifically, we find that, for any single MW polarization, at each reference sampling point $\vb{B}$ with $\abs{\vb{B}}<50~\unit{\micro\tesla}$, there are always directions in which small changes $\delta \vb{B}$ result in practically no change to the measurable signal. Since the signal does not change, it is impossible for an NN to determine from which exact magnetic field (i.e. $\vb{B}$ or $\vb{B}+\delta \vb{B}$) it was created. For different MW pulses, these problematic sweep directions may be different, allowing us to resolve this issue by using two MW pulses of different polarization in the main results. An example is shown in Figure \ref{fig:KLDisappereance1MW}. Note that the KL divergence will reduce for some directions in which the initial MW pulse performs well, due to the second MW pulse being less sensitive to these instead. \par
At higher magnetic field strengths $\abs{\vb{B}}>50~\unit{\micro\tesla}$, the problematic sweep directions still show small changes in the signal, thus making it possible to train the machine learning estimator. However, due to the reduced curvature in the KL divergence compared to the two MW polarizations, the sensitivity is degraded significantly, causing us to recommend the two MW polarization approach at any $\abs{\vb{B}}$.

\section{Alternate Experimental Scheme}\label{app:MethodB}
In the main text, we have focused on the results that yielded the highest accuracy and precision. However, 
achieving these does require that one is able to align the sensor with the known magnetic field quite well 
($\sim1~\unit{\micro\tesla}$, equivalent to about $0.5^\circ$ at $100~\unit{\micro\tesla}$). In order to avoid this constraint, we have tested the approach on a modified sampling region. \par 

Samples were drawn from a sphere that is placed in such a way that its center is at 
$\vb{B} = (0, 50, 21)^T\unit{\micro\tesla}$. It has a radius of $20~\unit{\micro\tesla}$ and thus contains a 
larger volume than the cylinder used to create the main results. We designed the sphere to cover the range of 
field strengths present in the background magnetic field of the Earth. This alternate method is most plausible
if one performs repeated measurements of only slightly varying $\vb{B}$ without the need for reorientation of 
the NV-based sensor. It is important to note that it is still necessary to ensure that $\vb{B}$ is within the 
restricted volume, such as by orienting the sensor based on an initial guess before the series of measurements. 
One could also ensure the initial $\vb{B}$ is at the center of the sphere by adding a constant 
magnetic field to offset the natural background, but this method should not be employed when zero-bias-field configurations are desired. Afterwards, given that $\vb{B}$ does not vary too significantly, 
one could track the changes in $\vb{B}$ using repeated measurements. 
To obtain results for the accuracy of such an approach, we have allowed for deviations of up to $20~\unit{\micro\tesla}$. Of course, as the modified sampling volume is more than 30 times larger than the approach in the main text, a minor increase in the standard deviation of 
estimates and the associated bias is expected. The bias may be reduced with more training samples or by reducing the volume if changes in $\vb{B}$ are small enough.
\begin{figure}
    \centering
    \includegraphics[width=1\linewidth]{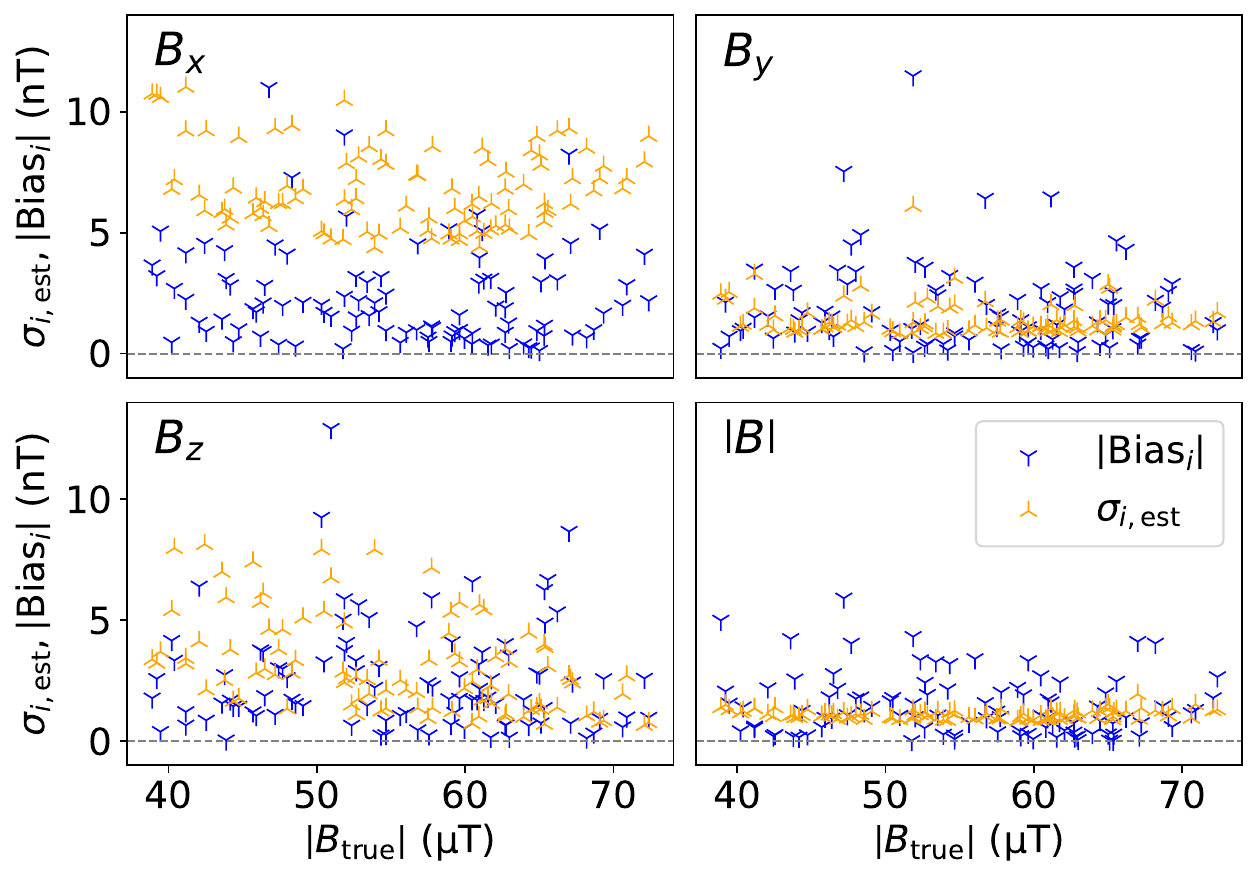}
    \caption{$\sigma_{i, \text{est}}$ and bias dependence on $\abs{\vb{B}_{\text{true}}}$ for the estimation with an SNR of $70\unit{\deci\bel}$ when sampling from a sphere.} 
    \label{fig:BiasSTDPointballbyBabs}
    %For some reason the placement of the later table bugs out and becomes weird if the figure isnt arranged with the full length of text in mind
\end{figure}
Results are depicted in Figure~\ref{fig:BiasSTDPointballbyBabs}. While the range of values is similar to those of 
our main results, we see significantly larger variations in the outcome for different test points. Otherwise, the 
main effect is a higher bias due to the larger sampling volume. It is possible to tune the volume size and thus 
bias with respect to the capabilities desired in experimental setups, otherwise the SNR ought to be decreased 
to ensure the accuracy of results. Alternatively, a combination of the NN estimator with an unbiased one, such 
as the KL divergence, may be fruitful. \par 
The variations between test points are likely due to the absence of strong correlations between the components 
of the true magnetic field, which in the previous data were a result of the narrow shape of the cylinder. Moreover, the weakest 
sampled magnetic field is now $\abs{\vb{B}_{\text{true}}}>35\unit{\micro\tesla}$, removing the range where the 
asymptotic behavior was clearly visible.  

\section{Overdetermined Estimation}\label{app:Overdetermined}
The NN-based estimation is used to determine the four parameters $B_x$, $B_y$, $B_z$, and $B_{\text{abs}}=\abs{\vb{B}}$. We have performed the estimation in Cartesian coordinates to enable easy comparison with other magnetometers. The estimation for $\abs{\vb{B}}$ is still included, as this parameter showed high estimation accuracy similar to $B_y$ and allowed simple comparison with previous work for the broadband MW magnetometry (\cite{afshar2025magnetometry}). The similarity to estimating $B_y$ is expected, as the axis $\vb{e}_r$ is closely aligned with $\vb{e}_y$ within the sampling region for $\vb{B}$. \par 
However, the NN output is overdetermined and, as we did not restrict the output parameters, one finds that $B_{x,\text{est}}^2 + B_{y,\text{est}}^2 + B_{z,\text{est}}^2 \neq \abs{\vb{B}}^2$, with minor deviations below the standard deviation of the estimates. Still, the additional estimate $\abs{\vb{B}}$ cannot be treated as a separate measurement, because it was generated from the same input data. \par
Nonetheless, for testing purposes, we have performed a maximum likelihood estimation (MLE) to ascertain the effect of including $\abs{\vb{B}}$ as a separate measurement. The maximum of the likelihood $L(\vb{B}_{\text{MLE}})$ was found by minimizing the right side of the equation 
\begin{equation}
\begin{split}
    -2\ln L(\vb{B}_{\text{MLE}}) = & (\vb{B}_{\text{MLE}} - \vb{B}_{\text{cart}})^\dag \vb*{\Sigma}_{\text{cart}}^{-1} (\vb{B}_{\text{MLE}} - \vb{B}_{\text{cart}}) \\ & + \frac{(\abs{\vb{B}}_{\text{MLE}} - \abs{\vb{B}})^2}{\sigma_{\text{est}}^2},
\end{split}
\end{equation}
where $\vb{B}_{\text{MLE}}$ is the MLE estimate, $\vb{B}_{\text{cart}} = (B_x, B_y, B_z)^T$, $\vb*{\Sigma}_{\text{cart}}$ is the corresponding covariance matrix for multiple noisy estimates of the same $\vb{B}_{\text{true}}$, and $\sigma_{\text{est}}$ is similarly the standard deviation for $\abs{\vb{B}}$. \par 
We find that the change to the estimate $\vb{B}_{\text{MLE}}-\vb{B}_{\text{cart}}$ is randomly distributed in each element. The difference is generally less than 10\% of the standard deviation for each $B_x$, $B_y$, $B_z$. This equates to a difference less than $1~\unit{\nano\tesla}$ for $B_x$ and less than $0.25~\unit{\nano\tesla}$ for $By$, $B_z$. As there is no consistent improvement and the change is relatively small, we recommend omitting this MLE step. Moreover, the higher estimation precision for $B_r$ cannot be used to improve the precision for $B_x$, because $\vb{e}_r$ is almost perpendicular to $\vb{e}_x$. \par
We have also tried calculating $\abs{\vb{B}}$ directly from the Cartesian estimate, instead of generating it as a separate output. This has a minimal effect on the results. While such an approach is preferable for a production-ready magnetometer to avoid inconsistencies in the estimate, one loses the ability to ascertain problems in the algorithm based on the separated $\abs{\vb{B}}$.

\section{Training Parameters}\label{app:NNParams}

We have used a feedforward NN (FNN) for our results. Other networks, such as neural ordinary differential 
equations (NODE) \cite{NEURIPS2018_69386f6b}, which yielded very good results on the one-dimensional problem 
presented in Ref.~\cite{afshar2025magnetometry}, were also investigated. With the setup described in this 
paper, the performance differences in the full vector magnetometry were minimal. For the sake of reduced 
training times and complexity of the approach, we have limited this paper to FNN.

\begin{table}
\setlength{\tabcolsep}{3pt}
\renewcommand{\arraystretch}{1.4}
\begin{tabular}{|c|c|} \hline
Parameter & Value \\ \hline
Dataset Size & 100000 \\
Loss & Mean Squared Error \\
Update Steps & 12100 \\
Learning Rate (LR) & $10^{-4.35}$ \\
LR Decay Factor & 0.1 \\
Steps each LR Decay & 3000 \\
Batch Size & 64 \\
NN Layers & 6 \\
Nodes per Layer & 512 \\
Activation & LeakyReLU \\ \hline
\end{tabular}
\caption{Hyperparameters used for training of main results. }
\label{tab:Hypparam}
\end{table}

A list of hyperparameters can be found in Table~\ref{tab:Hypparam}. Note that, for the method described in the main text, the first 3000 training steps 
were done on a cylinder with larger radius ($5~\unit{\micro\tesla}$), as this significantly improved initial convergence speed. 
Afterwards, the model was trained on the volume described in Sec.~\ref{subsec:SampleGen} for optimal results. 
Further parameters, such as those mainly used for regularization, have also been tested. All of them were 
removed for the main results as overfitting did not present any issues. We have balanced training time with 
final performance. With more resources, especially with larger datasets, improvements can be achieved. The 100000 samples were split 90/10 between training and validation. In actual experiments, 
one can use simulated data for most of the training and then fine-tune the model using experimental data. 
With the given parameters, training of a neural network could be completed within three hours using a current mid-range 
consumer-grade GPU.

\section{Bias}\label{app:bias}
Due to the finite training set giving limited information about the connection between $\vb{B}_{\text{true}}$ and the input $\vb{V}_{\text{in,ideal}}$ in an ideal system, there is a remaining estimation error $\Delta B$ in 
the NN-based estimation, which takes the form of a bias as
\begin{equation}
    \vb{B}_{\text{est,ideal}} = f_{NN}(\vb{V}_{\text{in,ideal}}) = \vb{B}_{\text{true}} + \Delta \vb{B}.
\end{equation}
The bias depends on the magnetic field $ \vb{B}_{\text{true}}$ that one tries to measure, and it is independent 
of the SNR. The mean for multiple independent measurements $N$ will converge to $\vb{B}_{\text{est,ideal}}$ 
and not $\vb{B}_{\text{true}}$.  \par 
Although we may not remove the bias entirely, we can reduce it so far that it becomes smaller than the 
uncertainty caused by the noise at a realistic SNR. Our two main methods are to keep the density of training 
samples as high as possible and to average $\vb{B}_{\text{est}}$ for independently trained neural networks. 
Their efficacy can be easily understood when we compare our problem to improving the accuracy of a fit to a 
function $\mathbb{R} \rightarrow \mathbb{R}$ from a set of finite data points. \par

Increasing the density of data points simply yields more information about the function we are trying to 
approximate, reducing our error for the approximation between data points.
By using multiple, independently trained NN, each will have a different $\Delta \vb{B}$. We can be almost 
certain that they will end up in different minima of the optimization landscape, because both the NN 
initialization and the sample order are randomized. Averaging over $\vb{B}_{\text{est}}$ gives us an average 
bias over all NN, which we can expect to be smaller than the $\Delta \vb{B}$ of a single, randomly picked NN. 
Still, for a large number of NN estimators, the average of all $\Delta \vb{B}$ does not converge to zero, as 
they were all trained using the same finite set of training samples. This training set might not capture all 
features of the function for determining $\vb{B}_{\text{true}}$. \par 
In our results, we have averaged the estimation over 20 NNs trained separately on the same dataset containing $N=100000$ data points. As we 
expect that future experiments will be limited in the number of samples they can generate, and the density of 
training samples $\vb{B}_k$ in a volume $V$ scales with $\frac{\sqrt[3]{N}}{V}$, we instead have elected 
to work with a small sampling volume, without sacrificing usability for the experimental approaches 
suggested in Sec.~\ref{subsec:SampleGen} and Appendix~\ref{app:MethodB}.

\end{document}